\documentclass[pre,twocolumn]{revtex4-1}

\usepackage{color}
\usepackage{enumerate}
\usepackage{amsfonts}
\usepackage{amssymb}
\usepackage{mathrsfs}
\usepackage[fleqn]{amsmath}
\usepackage[mathscr]{eucal}
\usepackage[english]{babel}
\usepackage{graphicx}

\newcommand{\comment}[1]{}
\newcommand{\var}{\mathrm{var}}
\newcommand{\rhoteq}{\rho_t- 1 - (2\epsilon - 2\pi\epsilon^2)k}
\begin{document}
\title{Analysis and Optimization of Population Annealing}

\author{Christopher Amey}
\email{camey@physics.umass.edu}
\affiliation{Department of Physics, University of Massachusetts,
Amherst, Massachusetts 01003 USA}

\author{Jonathan Machta}
\email{machta@physics.umass.edu}
\affiliation{Department of Physics, University of Massachusetts,
Amherst, Massachusetts 01003 USA}
\affiliation{Santa Fe Institute, 1399 Hyde Park Road, Santa Fe, New 
Mexico 87501 USA}

\begin{abstract}
Population annealing is an easily parallelizable sequential Monte Carlo algorithm that is well-suited for simulating the equilibrium properties of systems with rough free energy landscapes. In this work we seek to understand and improve the performance of population annealing.  We derive several useful relations between quantities that describe the performance of population annealing and use these relations to suggest methods to optimize the algorithm. These optimization methods were tested by performing large-scale simulations of the 3D Edwards-Anderson (Ising) spin glass and measuring several observables. The optimization methods were found to substantially decrease the amount of computational work necessary as compared to previously used, unoptimized versions of population annealing. We also obtain more accurate values of several important observables for the 3D Edwards-Anderson model.
\end{abstract}
\pacs{}
\maketitle

\section{Introduction}
Frustrated and disordered systems are ubiquitous in nature and consequently appear in a wide range of current research areas. Theoretical approaches to these systems are generally difficult to implement and have limited applicability and, as a result, computational methods are crucially important. Replica exchange Monte Carlo, also known as parallel tempering, \cite{SwWa86,Geyer91,marinari:92,HuNe96}  and multicanonical algorithms 
such as the Wang-Landau algorithm \cite{WaLa01,wang_landau}  have become the standard methods in the field because they partially overcome the problem of rough free energy landscapes. Perhaps unsurprisingly, the development of these algorithms is fundamentally intertwined with the frustrated systems they were invented to probe, and understanding the behavior of the algorithms contributes to understanding the system itself. 

One algorithm that is under active development is the population annealing Monte Carlo method. Population annealing (PA) is an example of a sequential Monte Carlo algorithm \cite{doucet:01} and was first introduced and applied to spin glasses by Hukushima and Iba \cite{hukushima:03}. Population annealing shares features with both parallel tempering and simulated annealing \cite{KiGeVe83}, however it has one major advantage: it is inherently parallel and can be easily implemented on large computing clusters.  Population annealing is closely related to nested sampling \cite{Skilling2006,Baldock2017}. 

Recently there has been a resurgence of interest in PA \cite{Mac10a}. There are several recent works that use PA to simulate not only spin glasses \cite{Wang2015,wang:15,Wenlong2016,Wenlong2017}, but also hard sphere fluids \cite{Callaham}, and the ferromagnetic Potts and Ising models using GPUs \cite{Weigel2017, Weigel2017-02}. In this work we continue the analysis of population annealing and derive several analytic results relating to its implementation. These results naturally lead to several optimization ideas that we test in the context of the 3D Edwards-Anderson spin glass.  Related optimization ideas are explored in \cite{BaPaWaKa17}.

In Sec.\ \ref{section_background}, we present the population annealing algorithm. We then introduce the Edwards-Anderson spin glass model and the relevant observables of interest, which is our test bed to study the optimization of PA. In Sec.\ \ref{section_theory} we introduce several observables and measures of error specific to population annealing. Furthermore, we derive an important inequality between measures of systematic and statistical errors, and we derive lower bounds for these errors in PA. We also introduce a method to determine an optimized annealing schedule. These relations suggest several improvements to PA, which we describe in detail in Sec.\ \ref{section_optimizations}. In Sec.\ \ref{section_results} we show the results of a large-scale simulation of the 3D Edwards-Anderson spin glass, which demonstrate our analytic results and the efficacy of the optimizations as well as providing new benchmark values of several quantities.

\section{Background}\label{section_background}
Population annealing (PA) is a sequential Monte Carlo algorithm designed to sample equilibrium states in systems with rough free energy landscapes. PA is similar to a parallel version of simulated annealing in many respects: a population of $R$ spin configurations, henceforth called ``replicas", is initialized at high temperature where equilibration is easy, and the population is slowly cooled. In addition, there is a resampling step at each temperature where individual replicas may be copied or eliminated according to their Boltzmann weights. Resampling ensures that the ensemble of replicas is kept near thermal equilibrium.  However, the resampled ensemble has degeneracies due to copied replicas, and does not fully sample the low-energy spectrum at the lower temperature. In order to address these issues, a Markov Chain Monte Carlo (MCMC) procedure such as the Metropolis algorithm is used to decorrelate and additionally equilibrate the population.

A PA simulation traverses an annealing schedule with $N_T$ inverse temperatures $\{\beta_{N_T-1},...,\beta_0\}$, where $\beta_{j+1}<\beta_j$, the initial temperature is typically infinite, $\beta_{N_T-1} = 0$, and the final $\beta_0$ is chosen as desired. Consider the resampling step from inverse temperature $\beta$ to $\beta'$. Each replica, denoted by a subscript $i$, is given a resampling factor $\tau_i$ that is proportional to the ratio of the Boltzmann weights of that replica at the two temperatures,
\begin{equation}
\label{eq:tau}
\tau_i = \frac{R}{\tilde{R}_\beta}\frac{e^{-(\beta'-\beta)E_i}}{Q(\beta,\beta')},
\end{equation}
where $E_i$ is the energy of replica $i$, $Q$ is a normalization factor used to control the population size,
\begin{equation}
\label{eq:Q}
Q(\beta,\beta') = \frac{1}{\tilde{R}_\beta}\sum_{i=1}^{\tilde{R}_\beta}e^{-(\beta'-\beta)E_i},
\end{equation}
and $\tilde{R}_\beta$ is the population size at step $\beta$. The actual population size $\tilde{R}_\beta$ may fluctuate around the target population size $R$.  The resampling factor is the expected number of copies of a replica, that is,  $\tau_i = \left<n_i\right>$, where $n_i$ is the stochastically chosen integer number of copies of a replica $i$. There are several ways to implement resampling in PA. In this work we follow the method used in Ref.\ \cite{Wang2015}, which minimizes the variance of $n_i$ by choosing $n_i(\beta,\beta')= \lfloor \tau_i(\beta,\beta') \rfloor$ with probability $\lceil \tau_i(\beta,\beta') \rceil - \tau_i(\beta,\beta')$ and $n_i(\beta,\beta')= \lceil \tau_i(\beta,\beta') \rceil$ otherwise. This method results in an average population equal to $R$, with fluctuations of order $\sqrt{R}$.  The resampling step ensures that, for large $R$, if the population is an equilibrium ensemble at $\beta$, it will also be an equilibrium ensemble at $\beta'$.

At the beginning of the simulation, each replica is statistically independent, however, after being copied, some replicas become statistically correlated with each other. Correlations are recorded via the ``ancestry'' of each replica. At the beginning of the simulation, each replica is labeled. During each resampling step, each replica passes its label to its offspring, and replicas that have the same ancestry label are said to be members of the same ``family''. Although the total number of families is always equal to the initial population size $R$, by the end of a simulation, most families will have no members. We shall see that the distribution of family sizes is closely related to both  statistical and systematic errors. \comment{cite PRE 92.0633...}

\subsection{Model and Observables}
\label{sec:modob}
We study and apply PA in the context of the 3D Edwards-Anderson (EA) spin glass, defined by the Hamiltonian
\begin{equation}
H = -\sum_{\left< n,m \right>}J_{nm} s_n s_m,
\end{equation} 
where the summation is over nearest neighbors on a cubic lattice with periodic boundary conditions, $s_{n}=\pm 1$ are Ising spins, and $J_{nm}$ are bonds drawn from a Gaussian distribution with zero mean and unit standard deviation. In order to understand the thermodynamic properties of the EA model, it is necessary to do many simulations with different bond configurations and to then take an average afterwards. PA gives access to the ensemble average of observables for a fixed set of bonds, therefore many PA simulations must be conducted to take a bond average. An average of an observable $O$ over bond configurations will be denoted as $[O]_\mathscr{J}$, whereas a thermal average will be denoted as $\left<O\right>$.

There are several types of observables that are theoretically interesting in spin glasses. Observables that can be measured directly in a single spin configuration are the easiest to measure since PA effectively simulates the canonical ensemble at each temperature step. Thermodynamic quantities for a single bond configuration such as the average energy $E=\langle H \rangle$ or average magnetization are all straightforward to measure as a simple population average at each temperature step. 

The order parameter for the EA model is obtained from the overlap between two spin configurations chosen independently from the canonical ensemble at a given temperature in the same bond configuration. The spin overlap $q$ is defined as
\begin{equation}
q = \frac{1}{N}\sum_{n=1}^{N} s_n^1 s_n^2,
\end{equation} 
where $s_n^1$ is the $n^{\rm th}$ spin of replica $1$, and $N$ is the total number of spins in the system. The thermal distribution of overlaps $P(q)$ for a given temperature and bond configuration has support on $\left[-1,1\right]$.  The Edwards-Anderson order parameter, $q_{\rm EA}$, is the thermal average of the absolute value of $q$ in (any) single pure thermodynamic state. In order to measure $P(q)$, it is necessary to measure $q$ many times from spin configurations $1$ and $2$  drawn independently from the equilibrium ensemble of replicas. This process is straightforward in PA as long as the family of each replica is recorded: PA gives access to the equilibrium ensemble, and replicas are guaranteed to be independent if they are from different families.

An important observable in spin glass theory is the disorder-averaged integrated overlap weight around the origin,
\begin{equation}
\mathrm{I}(q_0) = \left[\int_{-q_0}^{q_0} P(q) dq\right]_\mathscr{J},
\end{equation}
and following previous studies \cite{katzgraber:01,Wang2015}, we use $q_0=0.2$.  The quantity $\mathrm{I}(q_0)$ was introduced to distinguish between competing theories of the low temperature spin glass phase. Replica symmetry breaking theory \cite{parisi:79,parisi:80,parisi:83}  predicts that $\mathrm{I}(q_0)$ goes to a nonzero constant as $N \rightarrow \infty$ for any $q_0>0$ while the droplet  \cite{mcmillan:84, fisher:86, fisher:87, bray:85} and chaotic pairs pictures \cite{newman:92}  predict that $\mathrm{I}(q_0) \rightarrow 0$ as $N \rightarrow \infty$ for any $q_0< q_{\rm EA}$.

The link overlap $q_l$ is another quantity that is defined from two independent spin configurations,
\begin{equation}
q_l = \frac{1}{N_b} \sum_{\left< n,m \right>} s_n^1 s_m^1 s_n^2 s_m^2,
\end{equation}
where $N_b=3N$ is the number of bonds. Like the spin overlap, the link overlap is  useful for distinguishing theories of the spin glass phase \cite{katzgraber:01}.  It is also useful as a measure of equilibration using a relation found by Katzgraber and Young~\cite{katzgraber:01}.  They define
\begin{equation}
\Delta_{\rm KY} = [e_l - e]_\mathscr{J},
\end{equation}
where $e$ is the thermally averaged energy per spin and $e_l$ is an energy-like quantity defined from the link overlap,
\begin{equation}
e_l = -\frac{1}{T}\frac{N_b}{N}(1-\left< q_l \right>).
\end{equation}
For an individual bond configuration, it is not the case that $e_l = e$, however, for a disorder average over Gaussian bonds, $\Delta_{\rm KY}=0$. The requirement that $\Delta_{\rm KY} \approx 0$ is a useful indicator of equilibration. 

A free energy estimator, $\tilde{F}$, can be obtained \cite{Mac10a} at every step in the annealing schedule from the normalization factors [see Eq.\ (\ref{eq:Q})] of all previous annealing steps via
\begin{equation}
\label{eq:betaF}
-\beta_k \tilde{F}(\beta_k) = \sum_{l=k+1}^{N_T-1}\mathrm{ln}Q(\beta_l,\beta_{l-1})+\mathrm{ln}\Omega,
\end{equation}
where $\Omega=2^N$ is the number of microstates, $N_T$ is the total number of $\beta$ steps in the annealing schedule, and $\beta_l$ is the current step.\comment{cite PhysRevE.82.026704} 

If PA is run to a sufficiently low temperature we can obtain an estimator of the ground state energy $\tilde{E}_0$ from the lowest energy encountered during the simulation.  
Population annealing also gives direct access to the estimated probability of being in the ground state at each temperature, $\tilde{g}_0$, which is simply the fraction of the population at the lowest energy found during the entire simulation. Alternatively, with the aid of the free energy estimator, it is possible to obtain an indirect estimate of the probability of being in the ground state, $\bar{g}_0$, by calculating the Boltzmann weight,
\begin{equation}\label{eq:g0}
\bar{g}_0 = 2\,e^{-\beta E_0+\beta \tilde{F}}.
\end{equation}
By comparing the measured $\tilde{g}_0$ to the calculated $\bar{g}_0$, we can asses systematic errors.

\section{Population Annealing Theory}\label{section_theory}

\subsection{Error Estimation} \label{subsection_error}
As shown in \cite{Wang2015}, the systematic error of an observable $O$ is given by 
\begin{equation}
\Delta O = \mathrm{var}(\beta \tilde{F})\left[ \frac{\mathrm{cov}(\tilde{O},\beta \tilde{F})}{\mathrm{var}(\beta\tilde{F})} \right],
\end{equation}
where $\tilde{O}$ is the PA estimator of $O$ and the (co)variances are taken with respect to independent runs of PA. The bracketed quantity is expected to converge to a finite limit as $R\rightarrow\infty$, meaning that for large $R$, the systematic error for any observable is proportional $\mathrm{var}(\beta \tilde{F})$. Furthermore, the quantity $\mathrm{var}(\beta \tilde{F})$ is expected to scale as $1/R$, and so it is natural to define an equilibration population size, $\rho_f$, as
\begin{equation}
\label{eq:rhof}
\rho_f = \lim_{R\rightarrow\infty} R \hspace{1mm} \mathrm{var}(\beta \tilde{F}).
\end{equation}
The equilibration population size sets a population scale for a given bond configuration, and by choosing the population such that $R \gg \rho_f$,  systematic errors behave as  $\rho_f/R$. One complication is that in order to measure $\rho_f$, many simulations of the same bond configuration must be made.

An analogous quantity for statistical errors, $\rho_t$, can be defined that corresponds to the population size required to minimize statistical errors. In PA, if no decorrelating Markov chain Monte Carlo were done, the statistical errors would directly scale with the second moment of the family distribution, see \cite{Wang2015}. Therefore we define $\rho_t$ as
\begin{equation}
\rho_t = \lim_{R\rightarrow\infty} \frac{1}{R} \sum_{i=1}^R \eta_i^2, \label{rho_t_def2}
\end{equation}
where the summation is over families and $\eta_i$ is the size of family $i$.  Note that we can also express $\rho_t$ in terms of the variance of the family size distribution. Since the average family size is one,
\begin{equation}
\label{eq:defrhotvar}
\rho_t -1 = \lim_{R\rightarrow\infty} \mathrm{var}(\eta).
\end{equation}
Because we do perform MCMC at each annealing step, the actual statistical errors will be lower and $\rho_t$ can be used as a conservative estimate for the population size necessary to minimize statistical errors.  Specifically, the statistical error $\delta O$ in measuring an observable $O$ in a PA run with population size $R$ is bounded by
\begin{equation}
\delta O \leq  \sqrt{\frac{\mathrm{var}(O) \rho_t} {R}}
\end{equation}
where $\mathrm{var}(O)$ here refers to the underlying variance of the observable in the Gibbs distribution.

Unlike $\rho_f$, $\rho_t$ can be easily estimated from a single run and, as will be shown in the next section, there is a close relationship between $\rho_f$ and $\rho_t$ that can be used to our advantage. 

\subsection{Relation between $\rho_t$ and $\rho_f$}\label{section_theory_1}

In this section we provide an argument justifying the inequality
\begin{equation}
\rho_t -1 > \rho_f.
\end{equation}
The argument uses a modified version of PA where the exact free energy is known and is used for normalizing the weight of each spin configuration. This modified version of PA is similar to the idea of using blocks of a much larger total population to calculate $\rho_f$ and $\rho_t$. In this version of PA, the weight of spin configuration $i$ is given by
\begin{align}
\tau_i = e^{-(\beta' - \beta) E_i + \beta' F' - \beta F},
\end{align}
where $F$ and $F'$ are the exact free energies at $\beta$ and $\beta'$, respectively.  The actual number of copies $n_i$ of configuration $i$ is a random non-negative integer, $n(\tau_i)$, whose mean is $\tau_i$.  Our implementation of $n(\tau)$ is given in Sec.\ \ref{section_background} but the details are not important to the argument.  Here we divide $n_i$ into $\tau_i$ and a random remainder $z(\tau_i)$, whose mean is zero,
\begin{align}
n_i = \tau_i + z(\tau_i).
\end{align}
The total population $\tilde{R}_{\beta'}$ at temperature step $\beta'$ is given by
\begin{equation}
\label{eq:Rvn}
\tilde{R}_{\beta'} = \sum_{i=1}^{\tilde{R}_\beta} n_i,
\end{equation}
which can also be expressed in terms of the family size distribution,
\begin{equation}
\tilde{R}_{\beta'} = \sum_{j=1}^{R} \eta_j,
\end{equation}
where $\eta_j$ is the size of family $j$ at temperature step $\beta'$.
In the exact free energy version of PA,  families are independent of each other, so the variance of the sum is the sum of variances and $\mathrm{var}(\tilde{R}_{\beta'})= R \, \mathrm{var}(\eta)$.   From the definition of $\rho_t$, Eq.\ (\ref{eq:defrhotvar}),
\begin{equation}
\label{eq:rhotvarR}
\rho_t(\beta) - 1 =  \lim_{R\rightarrow\infty} \frac{1}{R} \, \mathrm{var}(\tilde{R}_\beta)
\end{equation}
Thus, in the exact free energy version of PA, there are relatively large fluctuations in the population size that scale as $\sqrt{\rho_t R}$ in contrast to the version we implement in the simulations, where the population size is nudged toward $R$ at every step and population fluctuations scale as $\sqrt{a R}$ with $a \lesssim 1$.

We can also derive an expression for $\rho_f$  that is related to population size fluctuations within the exact free energy version of PA.  Starting from the  factor $Q(\beta, \beta')$ [see Eq.\ (\ref{eq:Q})] we have,
\begin{align}
Q(\beta, \beta') &= \frac{1}{\tilde{R}_\beta} \sum_{i = 1}^{\tilde{R}_\beta} e^{-(\beta'-\beta) E_i} \nonumber \\
&= \frac{1}{\tilde{R}_\beta} \left( \sum_{i=1}^{\tilde{R}_\beta} \tau_i \right) e^{-\beta' F' + \beta F}.
\end{align}
We also know that $Q(\beta, \beta')$ is an estimator of the ratio of the partition functions of two subsequent temperatures so that,
\begin{equation}
Q(\beta, \beta') = e^{-\beta' \tilde{F}' + \beta \tilde{F}},
\end{equation}
where $\tilde{F}$ and $\tilde{F}'$ are the free energy estimators at $\beta$ and $\beta'$, respectively.
Combining the above two relations gives
\begin{align}
e^{-\beta' \Delta \tilde{F}' + \beta \Delta \tilde{F}} &= \frac{1}{\tilde{R}_\beta} \sum_{i=1}^{\tilde{R}_\beta} \tau_i \nonumber \\
&= \frac{1}{\tilde{R}_\beta} \sum_{i=1}^{\tilde{R}_\beta} [ n_i - z(\tau_i)],
\end{align}
where $\Delta \tilde{F} = \tilde{F} - F$ is the deviation of the free energy estimator from the exact free energy.
If the population is large, this deviation is small so we can expand the exponential of the free energy, and using Eq.\ (\ref{eq:Rvn}), we obtain
\begin{align}
1 - (\beta' \Delta \tilde{F}' - \beta \Delta \tilde{F}) = \frac{\tilde{R}_{\beta'}}{\tilde{R}_\beta} - \frac{\sum_{i = 1}^{\tilde{R}_\beta} z(\tau_i)}{\tilde{R}_\beta}.
\end{align}
The population size at temperature $\beta$ can be decomposed as $\tilde{R}_\beta = R + \delta \tilde{R}_\beta$, where $R$ is the mean population and $\delta \tilde{R}_\beta$ is the deviation from the mean at temperature $\beta$. Expanding in $\delta R/ R$ yields,
\begin{equation}
\begin{split}
1 - (\beta' \Delta \tilde{F}' - \beta \Delta \tilde{F}) =& \left( 1 + \frac{\delta \tilde{R}_{\beta'}}{R} \right) \left(1 - \frac{\delta \tilde{R}_\beta}{R}\right) \\
&- \frac{1}{R} \sum_{i = 1}^{\tilde{R}_\beta} z(\tau_i) \left( 1 - \frac{\delta \tilde{R}_\beta}{R} \right).
\end{split}\label{eq:deltaFbeta}
\end{equation}
From this point onwards, $\tilde{R}_j$ will denote the population at annealing step $j$ with inverse temperature $\beta_j$, and $\tau_i^j$ will denote the weight of configuration $i$ during the resampling step from $\beta_{j+1}$ to $\beta_j$. Disregarding all $(\delta R / R)^2$ terms, summing Eq.\ (\ref{eq:deltaFbeta}) over all steps in the annealing schedule, and taking the variance of the result yields,
\begin{equation}
\mathrm{var} \left( \beta_{k} \Delta \tilde{F}_k \right) = \mathrm{var} \left( \frac{\delta \tilde{R}_{k}}{R} - \frac{1}{R} \sum_{j = N_{T-1}}^{k + 1}\sum_{i = 1}^{\tilde{R}_{j+1}} z(\tau_i^{j})\right).
\end{equation}
From the definition of $\rho_f$, Eq.\ (\ref{eq:rhof}), we have,
\begin{equation}
\rho_f(\beta_k) =  \lim_{R\rightarrow\infty} \frac{1}{R} \mathrm{var} \left( \delta \tilde{R}_{k} - \sum_{j = N_{T-1}}^{k + 1} \sum_{i = 1}^{\tilde{R}_{j+1}} z(\tau_i^{j}) \right) .
\end{equation}
Expanding the variance and using Eq.\ (\ref{eq:rhotvarR}) yields a relation between $\rho_f$ and $\rho_t$,
\begin{equation}
\begin{split}
\rho_f(\beta_k) 
&= \rho_t(\beta_k) - 1  \\ 
 &- \lim_{R\rightarrow\infty} \frac{1}{R} \left[2\ \mathrm{cov} \left( \delta \tilde{R}_{k}, \sum_{j = N_{T-1}}^{k + 1} \sum_{i = 1}^{\tilde{R}_{j+1}} z(\tau_i^{j}) \right) \right.\vphantom{...} \\
&\left.\vphantom{...} -\mathrm{var} \left( \sum_{j = N_{T-1}}^{k + 1} \sum_{i = 1}^{\tilde{R}_{j+1}} z(\tau_i^{j}) \right)  \right].
\end{split}
\end{equation}
In Appendix \ref{appendix_cov} we argue that the quantity in square brackets is greater than zero yielding the desired inequality, $\rho_t -1 > \rho_f$.

There are two caveats concerning this inequality.  First, the argument in Appendix \ref{appendix_cov} establishing the positivity of the term in square brackets is not rigorous.  More importantly, the result applies to a version of PA that is normalized by the exact free energy and has large fluctuations in population size.  We conjecture that an ``equivalence of ensembles" result holds for the implemented and exact free energy version of PA so that both $\rho_f$ and $\rho_t$ are the same for both algorithms but this question deserves further study. 

We will see in the next two sections and in Appendix \ref{appendix_var} that the inequality between $\rho_f$ and $\rho_t - 1$ can be extended to an approximate equality, provided that the culling fraction is small at each step. This approximate equality and, by extension, the inequality are supported by numerical results shown in Sec.\ \ref{sec:rhofrhotnum}.

\subsection{Temperature step size, culling fraction and energy variance}
\label{section_theory_2}
As we shall see in Sec.\ \ref{sec:betastep}, a natural way  to choose the $\beta$--schedule for population annealing is to cull a fixed fraction of the population at each resampling step.  In this section we derive a relation between the culling fraction, the variance of the energy distribution, and the size of the temperature step. To derive this relation, note that the expected number of copies of each configuration is $\tau_i$ and the actual number of copies is $\lfloor \tau_i \rfloor$ with probability $\lceil \tau_i \rceil - \tau_i$ or $\lceil \tau_i \rceil$ otherwise. Thus a configuration can be eliminated only if $\tau < 1$, and the expected number of eliminated configurations in a resampling step is  
\begin{equation}
\label{eq_num_elim} 
\epsilon R = \sum_{\tau_i < 1} (1 - \tau_i), 
\end{equation}
where $\epsilon$ is the (expected) culling fraction. 
Let $\langle E \rangle$ and $\sigma_E^2$, respectively, be the thermal average energy and variance of the energy.  Consider a resampling step from $\beta$ to $\beta'$ with $\Delta\beta = (\beta'-\beta)$.  In the regime $\Delta\beta \sigma_E \ll 1$ we can expand the definition of $\tau_i$, Eq.\ (\ref{eq:tau}), to leading order in $\Delta \beta$ to obtain,
\begin{equation}
\tau_i = 1 - \Delta\beta (E_i - \langle E \rangle),
\end{equation}
meaning that $\tau$ is approximately a Gaussian random variable with mean one and standard deviation $\Delta\beta\sigma_E$.
Within this Gaussian approximation and for large $R$, the sum defining $\epsilon$ in Eq.\ (\ref{eq_num_elim})  can be replaced by an integral,
\begin{align}
\label{eq:eps}
\epsilon  & 
\approx  \int_{-\infty}^1 (1 - \tau)\mathscr{N}(\tau;1,\Delta\beta\sigma_E) \,\mathrm{d}\tau \nonumber\\
&=  \frac{\Delta\beta \sigma_E}{\sqrt{2\pi}},
\end{align}
where $\mathscr{N}(x;\mu,\sigma)$ is the pdf of the normal distribution with mean $\mu$ and standard deviation $\sigma$.
If we want to eliminate a fixed fraction of the population $\epsilon$, then the $\beta$--schedule must be chosen such that
\begin{equation}
\label{eq:cullbeta}
\Delta\beta \approx \frac{\epsilon \sqrt{2\pi}}{\sigma_E}.
\end{equation}

\subsection{Growth of $\rho_f$  in the MCMC-equilibrated regime}
\label{sec:rhofbound}
In the spin glass phase where the MCMC procedure alone is unable to equilibrate the system, the growth of $\rho_f$ depends, in a complicated way, on the structure and temperature-dependence of the free energy landscape.  However,  at high temperatures where the MCMC procedure is able to fully decorrelate replicas, which we refer to as the MCMC-equilibrated regime, we can show that $\rho_f$ is simply proportional to the number of annealing steps times the culling fraction.  
To understand the behavior of $\rho_f$ in the MCMC-equilibrated regime,  note that $\rho_f$ is defined in terms of the variance of $\beta F$,  [see Eq.\ (\ref{eq:rhof})], and take the variance of both sides of Eq.\ (\ref{eq:betaF}),
\begin{align}
\label{eq:varbfind}
\var[\beta_k \tilde{F}(\beta_k)] &= \var\left[\,\sum_{l=k+1}^{N_T-1}\mathrm{ln}Q(\beta_l,\beta_{l-1}) \right] \nonumber \\
&\approx \sum_{l=k+1}^{N_T-1}\var\left[\mathrm{ln}Q(\beta_l,\beta_{l-1})\right].
\end{align}
The second approximate equality becomes exact when the population is equilibrated by the MCMC procedure during each annealing step.  Using the definition of $Q$, Eq.\ (\ref{eq:Q}), and assuming the variation of $\tilde{R}$ is negligible, we expand $\var(\mathrm{ln}Q(\beta_l,\beta_{l-1}))$ to leading order in $(\Delta\beta_l) \sigma_E(\beta_l)$,
\begin{equation}
\label{eq:Qest}
\var\left[\mathrm{ln}Q(\beta_l,\beta_{l-1})\right] \approx \frac{1}{R} (\Delta\beta_l)^2 \sigma_E(\beta_l)^2,
\end{equation}
where $\sigma_E^2$ is the variance of the energy distribution.  Plugging Eq.\ (\ref{eq:Qest}) into Eq.\ (\ref{eq:varbfind}) yields, 
\begin{equation}
\var[\beta_k \tilde{F}(\beta_k)] = \sum_{l=k+1}^{N_T-1} \frac{1}{R} (\Delta\beta_l)^2 \sigma_E(\beta_l)^2.\label{eq:rho_f_est}
\end{equation}
From the relation between the size of the temperature step and the culling fraction, Eq.\ (\ref{eq:eps}), we find,
\begin{align}
R\,\var[\beta_k \tilde{F}(\beta_k)] = \sum_{l=k+1}^{N_T-1} 2\pi\epsilon(l)^2,
\end{align}
where $ \epsilon(l)$ is the culling fraction at the $l^{\rm th}$ annealing step.   Thus, for fixed culling fraction, $\epsilon$, we find that $\rho_f$ grows linearly in the number $k$ of annealing steps,
\begin{equation}
\label{eq:rhofgrow}
\rho_f = 2\pi\epsilon^2 k .
\end{equation}
This relation is valid if the culling fraction is small and enough MCMC sweeps are carried out in each annealing step that the replicas remain independent.
   
More generally, $\rho_f \geq 2\pi\epsilon^2 k$ and the inequality holds if the MCMC procedure is not able to keep the replicas fully decorrelated.

\subsection{Growth of $\rho_t$ in the MCMC-equilibrated regime} \label{sec_rhot_bound}
Similarly to the case of $\rho_f$, at high temperatures where the MCMC procedure is able to fully decorrelate the energy of the replicas at every annealing step, the growth of $\rho_t-1$ is  proportional to the number of annealing steps times the culling fraction.  To derive this relation we note that $\rho_t$ is equal to the variance of the family size distribution.  In the MCMC-equilibrated regime, the size of a given family is described by a birth and death process. In an approximation where the annealing step $k$ is taken to be a continuous ``time'' variable, the family size distribution, $P_\eta(k)$, is described by the Master Equation,
\begin{equation}
{\dot P}_\eta(k)= \epsilon \left[ (\eta-1)P_{\eta-1}(k) + (\eta+1)P_{\eta+1}(k) -2\eta P_\eta(k)   \right],
\end{equation}
where $\eta$ is the family size and $\epsilon$ is the culling fraction.  This is the classic birth--death process (see, for example, \cite{KrReBN10}) whose solution is 
\begin{equation}
\label{eq:petak}
P_\eta(k) = \left(\frac{1}{1 + \epsilon k}\right)\left(\frac{\epsilon k}{1 + \epsilon k}\right)^{\eta-1} \mbox{ for }  \eta\geq 1,
\end{equation}
and
\begin{equation}
P_0(k) = \left(\frac{\epsilon k}{1 + \epsilon k}\right) .
\end{equation}
From this distribution it is easily seen that 
\begin{equation}
\label{eq:rhotb}
\rho_t-1 = {\rm var}(\eta) = 2 \epsilon k.
\end{equation}
This equation holds in the MCMC-equilibrated regime where the number sweeps in each annealing step is greater than or comparable to the integrated autocorrelation time of the energy, so that the energy of every replica is independent of its family designation. 

Comparing Eqs.\ (\ref{eq:rhotb}) and (\ref{eq:rhofgrow}), we see that in the MCMC-equilibrated regime and for small culling fraction, 
\begin{equation}
\label{eq:rhofratiorhot}
\rho_f = \rhoteq .
\end{equation}
To a high level of accuracy, this relation also holds outside of the MCMC-equilibrated regime. This is shown numerically in Sec.\ \ref{sec:rhofrhotnum}, and supported analytically in Appendix \ref{appendix_var}.

\section{Optimization of Population Annealing}\label{section_optimizations}
In this work we focus on three general improvements to the population annealing algorithm: a method to choose the population size for each individual bond configuration, a way to choose the $\beta$--schedule to reduce statistical and systematic errors, and an ad hoc sweep schedule that improves equilibration.

\subsection{Hardness-dependent population size}
Previous authors have shown that the computational work required to equilibrate a specific bond configuration is approximately lognormally distributed for both population annealing and parallel tempering \cite{katzgraber:06,alvarez:10a-ea,Wang2015,YuKaMa13}. In population annealing, the work required to equilibrate a bond configuration is proportional to $\rho_f$, and so our first optimization takes advantage of the relation between $\rho_f$ and $\rho_t$ in order to tailor the population size necessary for each bond configuration, so we do not use too large a population on an easy system and spend resources inefficiently. In order to optimize the population for each bond configuration, it is usually necessary to do several simulations. An initial simulation is done with a small population, $R_0$.  From this simulation, we obtain an estimate of $\rho_t$, called $\rho_t(R_0)$.  If $R_0 > 100\rho_t(R_0)$ then we assume that $\rho_t(R_0) \approx \rho_t$ and the simulation was well-equilibrated. Otherwise we do another simulation with population size, $R_1 = 150\rho_t(R_0)$, which yields $\rho_t(R_1)$. This procedure is continued until the bond configuration is equilibrated according to the criterion of $R > 100\rho_t$. If $R_0$ is chosen wisely, this method converges quickly and uses far fewer resources than choosing a single $R$ adequate for all bond configurations. The values of $R_0$ used in our simulations are given in Table \ref{results:table1}.

\subsection{Optimal annealing schedule}
\label{sec:betastep}
In the MCMC-equilibrated regime it is possible to derive an optimal annealing schedule, which is composed of both the $\beta$--schedule and sweep schedule. This is done by minimizing $\rho_f$ while keeping the total amount of computational work, $W$, fixed. We define the total work as
\begin{align}
W &= \sum_{l=k+1}^{N_T-1} S(\beta_l) \\
&\approx \int_0^\beta \frac{S(\beta)}{\Delta\beta}d\beta,
\end{align}
where $S(\beta)$ is the sweep schedule, defined as the number of MCMC sweeps carried out per annealing step at inverse temperature $\beta$. 
In the MCMC-equilibrated regime we have an analytic expression for $\rho_f$ from Eq.\ (\ref{eq:rho_f_est}), which can also be approximated as an integral,
\begin{align}
\rho_f &= \sum_{l=k+1}^{N_T-1} (\Delta\beta_l)^2 \sigma_E(\beta_l)^2 \\
	&\approx \int_0^\beta \Delta\beta \sigma_E(\beta) d\beta.
\end{align}
We can use the method of Lagrange multipliers to minimize $\rho_f$ while holding $W$ constant by solving, 
\begin{equation}
0=\frac{\delta}{\delta \Delta\beta}(\rho_f + \lambda W),
\end{equation}
which yields, 
\begin{equation}
\sigma_E^2 - \lambda \frac{S(\beta)}{\Delta\beta^2} = 0.
\end{equation}
Equivalently,
\begin{equation}
\Delta\beta\sigma_E(\beta) \propto \sqrt{S(\beta)},
\end{equation}
and using Eq.\ (\ref{eq:cullbeta}),
\begin{equation}
\epsilon  \propto \sqrt{S(\beta)},\label{eq:constsweep}
\end{equation}
we see that in the MCMC-equilibrated regime, the optimal number of sweeps depends on the culling fraction at each step. In the case of a fixed culling fraction, a fixed sweep schedule is optimal.

Based on these ideas, in our simulations we employed a $\beta$--schedule that holds the culling fraction roughly constant. We observed that the resulting schedule does not depend strongly on disorder realization so we chose a schedule based on a single run and tested that the culling fraction was similar for several disorder configurations of varying difficulty. This schedule was then employed without modification for production runs.  Since, according to Eq.\  (\ref{eq:cullbeta}), $\Delta\beta = \epsilon \sqrt{2\pi}/{\sigma_E}$, the resulting schedule has many annealing steps at high temperature, where the standard deviation of the energy is large, and few annealing steps at low temperature. A similar uniform $\epsilon$ scheme was used in PA simulations of hard spheres \cite{Callaham}.  Since the variance of the energy scales linearly with the system size (except near the critical point), the same $\beta$--schedule can be used for many system sizes, though for larger sizes the uniform culling fraction will increase.  If the range of sizes studied is too large, interpolating temperatures can be added to the schedule to reduce the culling fraction. Although it is not theoretically well-justified in the critical or the low temperature regimes, we continued to use a uniform culling fraction to determine the $\beta$--schedule over the entire temperature range of our simulations. 

In accordance with Eq.\ (\ref{eq:constsweep}), we chose a sweep schedule that was fixed in the high temperature regime. We found that focusing most of the computational work in the glass transition regime minimized $\rho_t$, and that in the glassy phase it was favorable to rely on PA resampling and do very little MCMC work. Our ad hoc sweep schedule had three Monte Carlo sweeps performed for $\beta < 0.5$, 22 for $0.5 \leq \beta \leq 2.5$, and a single sweep for $\beta > 2.5$.  Our annealing schedule is not directly comparable with the annealing schedules studied in Ref.\ \cite{BaPaWaKa17}.  While our $\beta$-schedules is the same as their ``std($E$)'' schedule, the computationally meaningful density of sweeps per unit $\beta$ is not the same as for any of their annealing schedules because of our non-uniform sweep schedule. A deeper understanding of the optimal annealing schedule over the whole range of temperatures remains an open problem.

\section{Results}\label{section_results}
\label{sec:results}

\begin{table}
	\begin{center}
		\begin{tabular}{| l | c | c | c |}
			\hline
			& L=6 & L=8 & L=10 \\ \hline
			Culling fraction, $\epsilon$ & $0.103$ & $0.168$ & $0.272$ \\ \hline
			Sweeps per replica & $1005$ & $1005$ & $1009$ \\ \hline
			Temperature steps, $N_T$ & $95$ & $95$ & $99$ \\ \hline
			Initial population, $R_0$& $2\times10^3$ & $2\times10^3$ & $2\times10^4$ \\ \hline
			Maximum population & $2.64\times10^5$ & $1.5\times10^7$ & $1.5\times10^7$ \\ \hline
            Disorder samples& $2\times10^4$ & $2\times10^4$ & $5\times10^3$ \\ \hline
			Unequilibrated samples & 0 & 4 & 145 \\ \hline
		\end{tabular}
		\end{center}
 		\caption{Parameters of the simulations for each of the three system sizes. 
        }
		\label{results:table1}
\end{table}
This section contains two types of numerical results. Sections \ref{sec:results} A-D address the behavior of PA, validating theoretical predictions and testing the optimization ideas described in previous sections. Section \ref{sec:results} E  reports measurements from large-scale simulations of several observables for the 3D Edwards-Anderson spin glass in order to compare with previous work and improve the state-of-the-art. The parameters used in this work are shown in Table\ \ref{results:table1}.

\subsection{Relationship of $\rho_f$ and $\rho_t$}
\label{sec:rhofrhotnum}
\begin{figure}
	\hspace{-6mm}\includegraphics[width=0.45\textwidth]{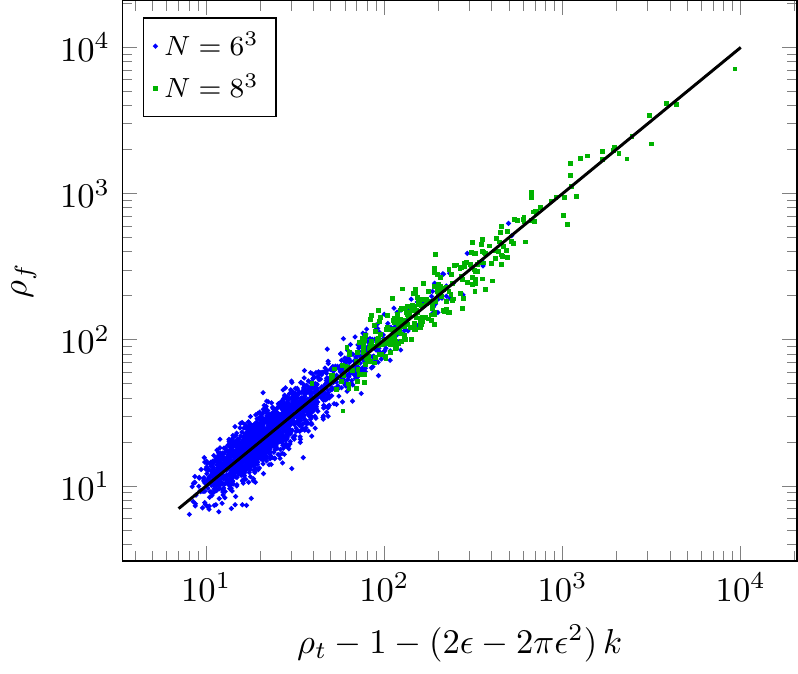}
	\caption{Scatter plot of $\rhoteq$ vs $\rho_f$ at $\beta=5$ ($k=95$) for $L=6$  and $L=8$; each data point corresponds to a single bond configuration. The solid line corresponds to  $\rho_t-1 + (2\pi\epsilon^2 - 2\epsilon)k=\rho_f$, see Eq.\ (\ref{eq:rhofratiorhot}). 
    }
	\label{results:fig1}
\end{figure}

In Sec.\ \ref{section_theory_1} we showed that $\rho_t-1 \geq \rho_f$ and, in Eq.\ (\ref{eq:rhofratiorhot}), gave an approximate relation between these two quantities. Here we test these relationships. Our proposal to optimize population size for each disorder realization relies on the easily measured $\rho_t$ as a proxy for the more difficult to measure $\rho_f$, so it is important to determine the relationship between these two quantities. 

In order to accurately measure $\rho_f = R \, \var(\beta \tilde{F})$, we ran population annealing 48 times for each configuration, with population sizes chosen such that $R \geq 100\rho_t$. This ensured that each simulation was well-equilibrated and that we could measure $\var(\beta\tilde{F})$ with reasonable accuracy.  We calculated $\rho_f$ for 2000 $L=6$ samples and for 300 $L=8$ samples. Calculating the error of $\rho_f$ is equivalent to calculating the error of a sample variance. As shown in previous work \cite{Wang2015}, $\beta \tilde{F}$ taken from a well-equilibrated bond configuration is normally distributed, which makes estimation of the error of $\var(\beta \tilde{F})$ particularly easy \cite{wilks:62},
\begin{equation} \label{gaussian_err}
	\var\left[\var(\beta \tilde{F})\right] = \frac{2}{M-1}\var(\beta \tilde{F})^2,
\end{equation}
where $M$ is the number of trials. The corresponding error in $\rho_f$ is
\begin{equation}
	\delta\rho_f = \sqrt{\frac{2}{M-1}} \rho_f,
\end{equation}
which for 48 trials gives a relative error, $\delta\rho_f/\rho_f$, of about 21\%. Since $\rho_t$ is calculated from a single disorder realization, it is expected that $\delta\rho_t \ll \delta\rho_f$. We find this to be true empirically, with $\delta\rho_f \approx 20 \delta\rho_t$ for $L=6$, and $\delta\rho_f \approx 16 \delta\rho_t$ for $L=8$ simulations. 

Figure \ref{results:fig1} is a scatter plot of $\rhoteq$ vs $\rho_f$ at $\beta=5$, where each point corresponds to one disorder realization and the solid line corresponds to $\rho_f = \rhoteq$.  The results are consistent with $\rho_f \approx \rhoteq$ holding for all disorder realizations.

\begin{figure}
	\hspace{-6mm}\includegraphics[width=0.43\textwidth]{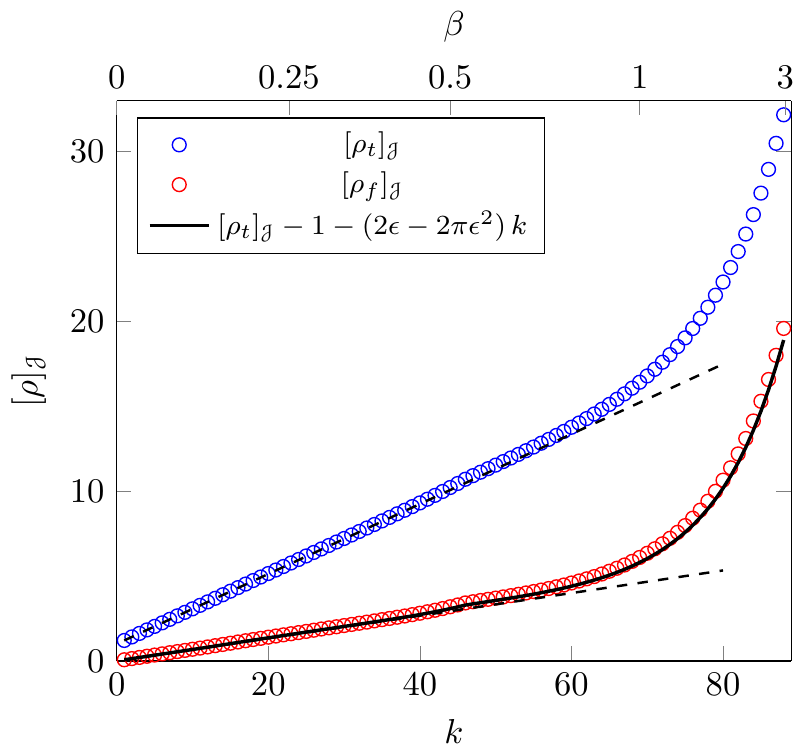}
    \caption{$[\rho_f]_\mathscr{J}$ and $[\rho_t]_\mathscr{J}$ of $L=6$ as a function of annealing step $k$, with the nonlinear $\beta$ scale on the upper x-axis. Dashed lines correspond to the theoretically predicted MCMC-equilibrated estimates, Eqs.\ (\ref{eq:rhofgrow}) and (\ref{eq:rhotb}). The solid line corresponds to the difference $[\rho_t]_\mathscr{J}-1-(2\epsilon-2\pi\epsilon^2)\,k$,  see Eq.\ (\ref{eq:rhofratiorhot}).}
	\label{results:fig2}
\end{figure}

In the MCMC-equilibrated regime, $\rho_f = 2\pi\epsilon^2 k$ and $\rho_t = 1 + 2 \epsilon k$ as shown in Secs.\ \ref{sec:rhofbound} and \ref{sec_rhot_bound}, respectively. Figure \ref{results:fig2} shows $[\rho_f]_\mathscr{J}$ and $[\rho_t]_\mathscr{J}$ for $L=6$ with the dashed lines representing the theoretical linear dependence on number of annealing steps $k$. The solid line represents the estimated value of $\rho_f$ calculated using $[\rho_t]_\mathscr{J}-1-(2\epsilon-2\pi\epsilon^2)\,k$. The estimated value was found to be within 5\% of the true value of $\rho_f$ for all annealing steps for both $L=6$ and $L=8$ (not shown).  Note that the sharp rise in both $\rho_f$ and $\rho_t$ occurs near the critical temperature, $\beta_c = 1.05$.

\begin{figure}
	\includegraphics[width=0.46\textwidth]{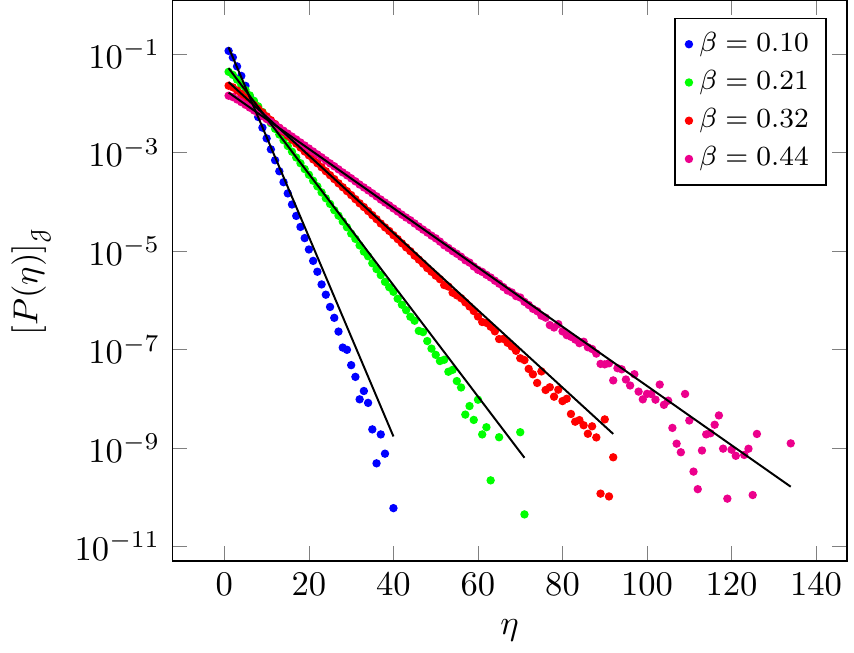}
	\caption{The disordered-averaged family size distribution $[P(\eta)]_\mathscr{J}$ as a function of family size $\eta$,  at several temperatures all in the high temperature regime (the $\beta$ of each distribution increases from left to right). The distributions are exponential and have shape parameters that match the predictions of Eq.\ (\ref{eq:petak}). The value at $\eta=0$ is not shown in this plot.}
	\label{results:fig2b}
\end{figure}

Figure \ref{results:fig2b} shows the disorder-averaged family size distribution for several temperatures in the MCMC-equilibrated regime for $L=8$ and confirms the prediction of Sec.\ \ref{sec_rhot_bound} of an exponential family size distribution.  The straight lines in the figure are obtained from Eq.\ (\ref{eq:petak}) and show that there is good quantitative agreement except for tail of the distribution. The higher values of $\eta$ are underrepresented, especially for low values of $\beta$, due to the finite size of the population.

\subsection{Distribution of $\rho_t$}

\begin{figure}
	\hspace{-6mm}\includegraphics[width=0.5\textwidth]{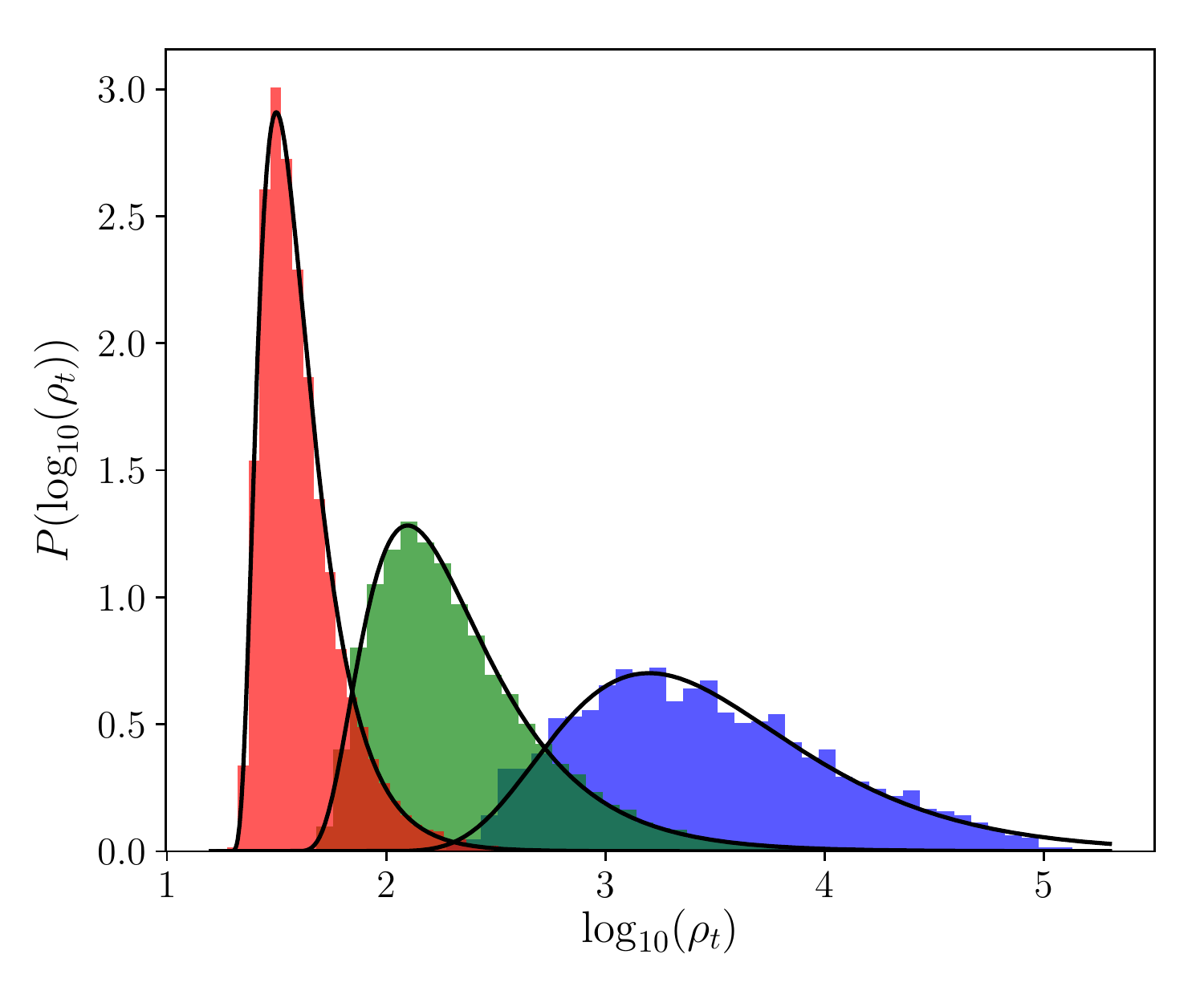}

	\caption{The distribution of $\log_{10} \rho_t$ for system sizes $L=6$ (left), 8 (middle) and 10 (right).
The solid lines are inverse Gaussian fits with the parameters given in Table \ref{results:inv_gauss_tbl}. }

	\label{results:fig5}
\end{figure}

It is known  \cite{alder:04,katzgraber:06,alvarez:10a-ea,YuKaMa13,Wang2015}  that the computational hardness of simulating  Ising spin glasses has a broad distribution with respect to disorder realizations.  In the context of a MCMC algorithm such as parallel tempering the computational hardness is typically measured by the exponential or integrated auto-correlation times.  These quantities have been found to be approximately lognormally distributed.  For population annealing, we may use $\rho_f$ or $\rho_t$ for the purpose of measuring computational hardness.  Figure \ref{results:fig5} shows histograms of $\log_{10} \rho_t$ for the three system sizes simulated.  We found that a log inverse Gaussian distribution is an excellent fit to the $\rho_t$ distributions.  The fits are shown as solid lines in the figure.  The three-parameter inverse Gaussian distribution is defined by,
\begin{equation}
\label{eq:invgauss}
P(x; \mu, \lambda,  l) = \sqrt{\frac{\lambda}{2\pi (x-l)^3}}\mathrm{exp}\left[\frac{-\lambda(x- \mu - l)^2}{2\mu^2(x- l)} \right]
\end{equation}
with $x=\ln \rho_t$. The parameters of the fits are given in Table\ \ref{results:inv_gauss_tbl}. The shift parameter $l$, which shifts the support of the distribution from $(0,\infty)$ to $(l,\infty)$, is necessary since $\rho_t$ is bounded away from zero. A rough estimate of shift parameter can be obtained by assuming that the easiest bond realizations are  MCMC-equilibrated for all temperatures, resulting in $l=\log(\rho_t^{\mathrm{min}}) \approx \log(1 + 2\epsilon k_{\mathrm{max}})$. The values for $l$ obtained from this formula are $2.97$, $3.50$ and $4.00$ for sizes $L=6$, 8 and 10, respectively.  These values are in reasonable agreement with the fitted values shown in Table \ref{results:inv_gauss_tbl}.  The log inverse Gaussian also works well to fit the $\rho_f$ distribution with a shift parameter predicted by $\rho_f^{\mathrm{min}}=2\pi\epsilon^2k$.

We can also compare the disorder averaged value of $\rho_t$ with the predictions from the fit.  The mean of $\rho_t=e^x$ is given by,
\begin{equation}
[\rho_t]_{\mathscr{J}} = \exp\left[ l + \frac{\lambda}{\mu}( 1- \sqrt{1-2 \mu^2/\lambda} ) \right],
\end{equation}
for $\lambda/2 \mu^2 >1$.  The tail of the inverse Gaussian is exponential, so if $\lambda/2 \mu^2 \leq 1$, the mean of $e^x$ is infinite.   The fitted values of $[\rho_t]_{\mathscr{J}}$ obtained from this equation are shown in  Table \ref{results:inv_gauss_tbl}.  The values computed directly from the data are $[\rho_t]_{\mathscr{J}}=46$, 385 and 12700 for $L=6$, 8 and 10, respectively. The large discrepancy between the  $L=10$ fitted and measured values may be due to several factors.  First, the number of samples for $L=10$ is smaller than for $L=6$ and 8, so the tail of the $\rho_t$ distribution may not be fully  sampled.  Second, a significant fraction of $L=10$ samples were not equilibrated and, for these samples, we have most likely underestimated $\rho_t$. If the tail of the distribution is properly described by the log inverse Gaussian, the value of $[\rho_t]_{\mathscr{J}}$ obtained from the fit may be more accurate than the average of the $\rho_t$ data from a finite sample size. On the other hand, for $L=10$, the ratio $\lambda/2 \mu^2 =1.04$, which is quite near the divergence at $\lambda/2 \mu^2 =1$ so results for $[\rho_t]_{\mathscr{J}}$ may be highly sensitive to errors in the fit.  The near divergence of $[\rho_t]_{\mathscr{J}}$ for $L=10$ also suggests that the annealing schedule for this size should have either more temperature steps or more sweeps per step. 

\begin{table} 
	\begin{center}
		\begin{tabular}{| l | c | c | c |}
			\hline
			 & L=6 & L=8 & L=10 \\ \hline
			$\mu$ & 0.83(1) & 1.97(2) & 4.53(13) \\ \hline
			$\lambda$ & 3.25(16)  & 9.28(35) & 42.7(40) \\ \hline
			$ l$ & 2.88(1) & 3.39(2) & 3.50(13) \\ \hline
            $[\rho_t]_{\mathscr{J}}$& 46.4 & 491 & 66500 \\ \hline
		\end{tabular}
		\end{center}
		\caption{Fits of the $\rho_t$ distribution to a log inverse Gaussian distribution defined in Eq.\ (\ref{eq:invgauss}).} 
		\label{results:inv_gauss_tbl}
\end{table}

As seen in Table \ref{results:inv_gauss_tbl}, the computational effort required to reach equilibrium scales up rapidly with system size and is broadly distributed.  How does this effort translate into wall clock time on a modern CPU?  For $L=10$ the algorithm's run time on a single CPU is approximately 0.03 sec/replica.  The typical value of $\rho_t$ for $L=10$, defined by $\exp([\log \rho_t]_{\mathscr{J}})$, is approximately 3000 and the equilibration criterion is that $R \geq 100 \rho_t$ thus the typical running time is approximately 2.5 hours. This relatively benign number is, however, misleading because of the exponential tail of the $\log \rho_t$ distribution.  If one instead takes the average hardness, $[\rho_t]_{\mathscr{J}}\approx 66500$, predicted from the inverse Gaussian fit (see Table \ref{results:inv_gauss_tbl}), then the average running time to equilibrate every disorder realization in a very large $L=10$ sample would be approximately 55 hours per disorder realization.  This number exceeds the computing time expended on our $L=10$ simulations since we did not equilibrate all disorder realizations.  Of course, the algorithm can be efficiently parallelized so the wall clock time per replica can be made much smaller than these numbers.

\subsection{Optimized vs.\ unoptimized annealing schedule}

In this section we compare the performance of an optimized and unoptimized annealing schedule used in PA.  Our optimized annealing schedule has a $\beta$--schedule that keeps the culling fraction $\epsilon$ nearly constant, and an ad hoc sweep schedule that concentrates sweeps over a range of temperatures around the critical point, as  described in Sec.\ \ref{sec:betastep}.  The unoptimized annealing schedule has constant $\beta$ steps, $\Delta \beta = 0.05$, with 10 sweeps per step, and is similar to the annealing schedule used in Ref.\ \cite{Wang2015}.

The  figure of merit that we wish to minimize is size of the systematic errors, which scale as $\mathrm{var}(\beta F)$, times the total computational work, $W=RS$, where $R$ is the population size and $S$ is the total number of Monte Carlo sweeps per replica.  Using the result  $\rho_f \approx \rho_t$ we have that 
\begin{equation}
	W \, \mathrm{var}(\beta F) \approx S \rho_t.
\end{equation}
We have used the same number of sweeps in both the optimized and unoptimized algorithms so the comparison of the performance of the algorithms reduces to the comparison of $\rho_t$.

\begin{figure}
	\hspace{-6mm}\includegraphics[width=0.45\textwidth]{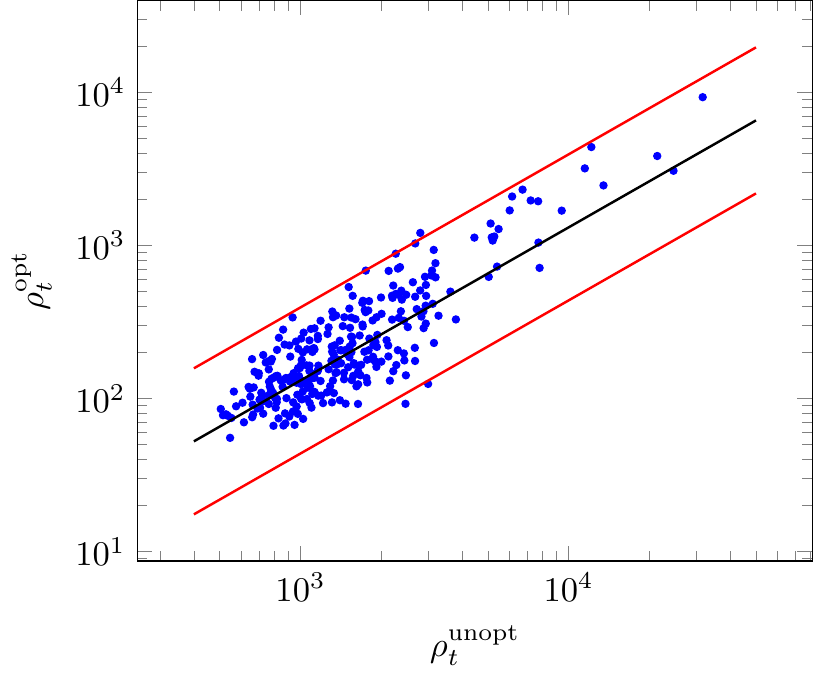}
	\caption{Values of $\rho_t$ at $\beta=5$ for the optimized and unoptimized annealing schedules. Each point corresponds to one of 300 $L=8$ bond configurations. The horizontal coordinate of each point is the unoptimized value and the vertical coordinate the optimized value of $\rho_t$.
The central (black) line corresponds to an improvement of the optimized relative to the unoptimized annealing schedule  by a factor of 7.6, and the upper and lower (red) lines correspond to factors of $7.6/3$ and $7.6\times3$, respectively.}
	\label{results:fig3}
\end{figure}

Figure \ref{results:fig3} is a scatter plot of the values of $\rho_t$, the vertical position of each point is the optimized $\rho_t^{\mathrm{opt}}$, and the horizontal position, $\rho_t^{\mathrm{unopt}}$, the unoptimized value. Each point represents one of 300 disorder realizations for system size $N=8^3$ at $\beta=5$.  The plot shows that the unoptimized algorithm is, on average, less efficient by a factor of 7.6. This means that, on average, the unoptimized algorithm requires a population 7.6 times larger to achieve the same quality of results as the optimized algorithm. 

\subsection{$L=6,8,10$ results}

\begin{figure}
	\hbox{\hspace{-5mm}\includegraphics[width=0.56\textwidth]{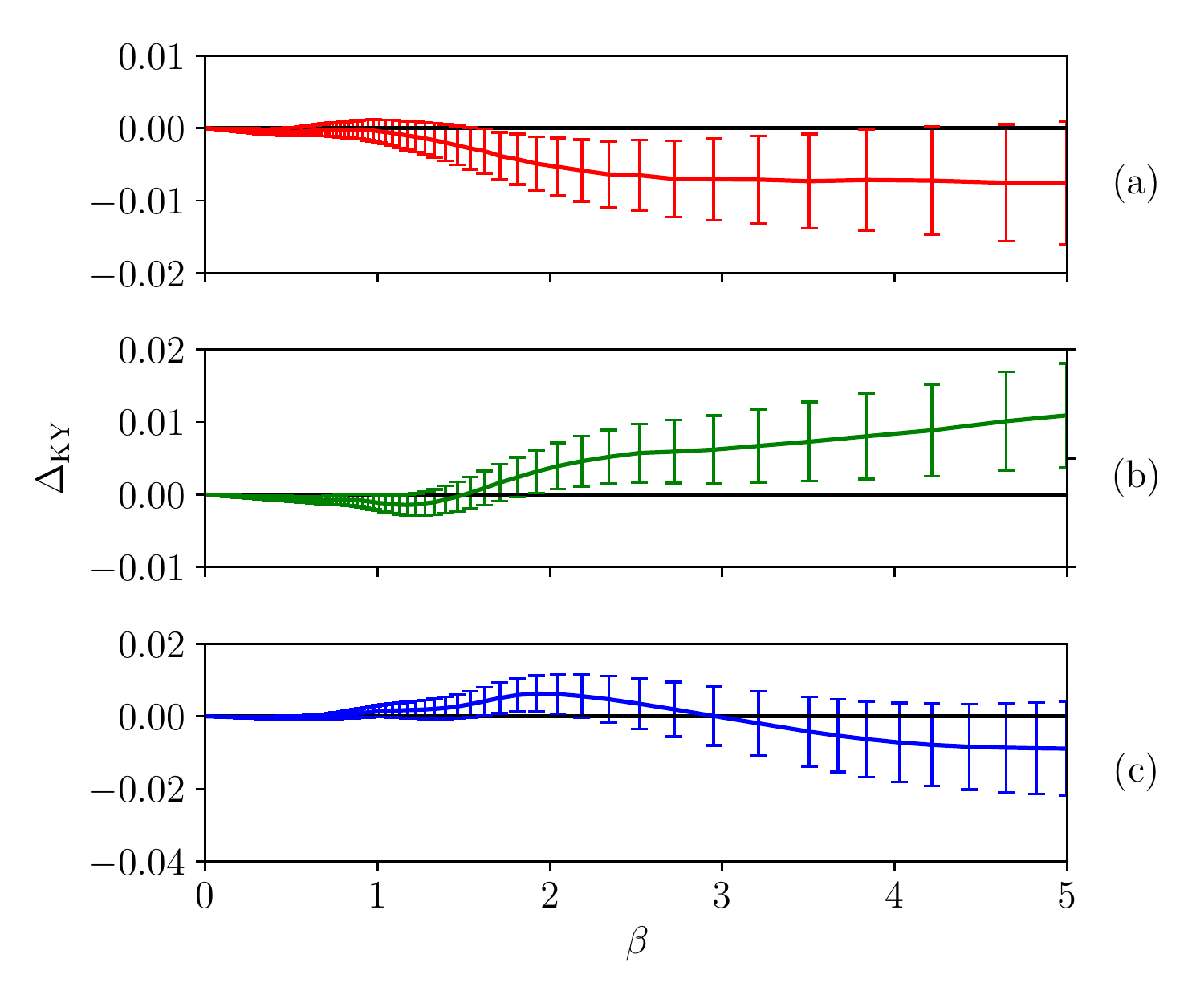}}
	\caption{$\Delta_{\rm KY}$ as a function of inverse temperature $\beta$ for $L=6$ (a), $L=8$ (b), and $L=10$ (c).
    }
	\label{results:fig4}
\end{figure}

\begin{table}
	\begin{center}
		\begin{tabular}{| l | c | c | c |}
			\hline
			& L=6 & L=8 & L=10 \\ \hline
			$\mathrm{I}(0.2)$ & 0.0188(5) & 0.0185(5) & 0.0185(10) \\ \hline
			$\Delta_{\rm KY}$ & $-0.0075(85)$& 0.010(7) & $-0.009(13)$\\ \hline
			$\left[E_0/N\right]_\mathscr{J}$ & $-1.6891(4)$ & $-1.6951(2)$ & $-1.6976(3)$ \\ \hline
		\end{tabular}
		\end{center}
		\caption{The integrated overlap $\mathrm{I}(0.2)$, Katzgraber-Young equilibration measure $\Delta_{\mathrm{KY}}$, and the bond-averaged ground state energy per spin $\left[E_0/N\right]_\mathscr{J}$, all measured at $\beta=5$. 
        }
		\label{results:table2}
	
\end{table}

\begin{table}
	\begin{center}
		\begin{tabular}{| l | c | c |}
			\hline
			L &\hspace{2mm} $[\log_{10}\tilde{g}_0]_{\mathscr{J}}$ \hspace{2mm} & $\log_{10}2-\beta [(\tilde{E}_0-\tilde{F}) ]_{\mathscr{J}}/\mathrm{ln}(10)$ \\ \hline
			6 & $-$0.7577(23) & $-$0.7549(23) \\ \hline
			8 & $-$1.6957(34) & $-$1.6900(34) \\ \hline
			10 & $-$3.2251(93) & $-$3.2104(92) \\ \hline
		\end{tabular}
		\end{center}
		\caption{Comparison of the disorder average of the logarithm of the fraction in the ground state, $[\log_{10}(\tilde{g}_0)]_{\mathscr{J}}$ at $\beta=5$ to the indirect measure, $[\log_{10}(\bar{g}_0)]_{\mathscr{J}}$ based on the Boltzmann factor, see Eq.\ (\ref{eq:g0}).}
		\label{results:table3}	
\end{table}

To test the optimized algorithm and obtain state-of-the-art results for observables described in Sec.\ \ref{sec:modob}, we ran large-scale simulations of the 3D EA spin glass for three system sizes, with parameters provided in Table \ref{results:table1}. 
As seen in this table, the equilibration standard, $R \geq 100\hspace{1mm}\rho_t$, was met by nearly all configurations for $L=6$ and $L=8$, but approximately 3\% of the configurations for $L=10$ remained unequilibrated when the maximum population was restricted to $R=1.5\times10^7$. The equilibration standard used was higher than previous papers employing PA to study spin glasses, and some systems that we rejected as unequilibrated would have been accepted previously.

It is worth emphasizing that the adaptive population scheme allowed us to sample more bond configurations than most previous studies, while ensuring that nearly all configurations were well-equilibrated. As a result, errors associated with a finite number of bond configurations are especially low for $L=6$ and 8, where we used $2\times10^4$ samples. The statistical errors reported in Table \ref{results:table2} and \ref{results:table3} are obtained from the standard deviation of the observable with respect to disorder realization and do not include errors associated with individual disorder realizations.   Despite the large number of disorder realizations, the error due to the finite sample size is substantially larger than the contribution from systematic and statistical errors of each disorder realization, as shown in \cite{Wang2015}. 

Overall, the results in Table \ref{results:table2} are consistent with those found in previous works \cite{wang:15, Wang2015}. The average ground state energy, $\left[E_0\right]_\mathscr{J}$, is within error bars of previous measurements and, as shown in Fig.\ \ref{results:fig4}, $\Delta_{\rm KY}$ is close to zero for all $\beta$, as which is consistent with a well-equilibrated set of samples. Our values of $\mathrm{I}(0.2)$ are all within two standard deviations of those found in Ref.\ \cite{Wang2015}, however, it is noteworthy that our values are consistently lower. Despite the slight difference in value, the trend that $\mathrm{I}(0.2)$ remains constant over several system sizes is evident.

Table \ref{results:table3} shows the disorder average of the logarithm of the directly measured fraction in the ground state $\tilde{g}_0$ and the indirectly measured quantity  $\bar{g}_0$ calculated using the Gibbs distribution, Eq.\ (\ref{eq:g0}). Although the two methods yield values that are within error bars, the computed $\bar{g}_0$ appears to be consistently larger than the measured $g_0$.  To leading order, the free energy estimator is systematically larger than the actual free energy by $F = \tilde{F} + \rho_f/2 \beta R$  \cite{Wang2015}, so it is expected that $\bar{g}_0$ would be systematically larger than $g_0$. It should be noted that by definition $g_0$ cannot be zero because the ground state energy is here defined as the lowest energy replica found,  even if this is not the true ground state.  

\section{Conclusions}\label{section_conclusion}

This paper makes several contributions to understanding and improving the population annealing algorithm, especially as applied to spin glasses:  we have studied the behavior of two important measures of equilibration for population annealing, optimized the algorithm in several ways, and obtained state-of-the-art results for several important spin glass observables.  Our results help put population annealing on a firmer footing as an effective tool for highly parallelized simulations of disordered systems such as spin glasses that have rough free energy landscapes.  While this paper focuses on the three-dimensional Edwards-Anderson model, many of the theoretical results and optimization methods are applicable to population annealing simulations of a much broader class of systems. 

The two equilibration measures, $\rho_t$ and $\rho_f$, set the population size needed to control statistical and systematic errors, respectively.  The equilibrium population size $\rho_f$ is based on the variance of the free energy and is the more fundamental measure of systematic errors but more difficult to accurately measure in a single simulation.  We have demonstrated that $\rho_f \leq \rho_t-1$, and confirmed that these two quantities are close to being equal when both are large.  We have also shown that in the MCMC-equilibrated regime, $\rho_f$ and $\rho_t$ each grow linearly in the number of annealing steps, a fact that can be used to design optimal annealing schedules.  Finally, we have shown that for the 3D EA spin glass, the distribution of $\log \rho_t$ values is accurately described by an inverse Gaussian distribution.

We have shown that there are a number of simple modifications which improve the efficiency of population annealing.  We have also shown that a $\beta$--schedule that is chosen by fixing the culling fraction is optimal in the MCMC-equilibrated regime. Lastly, we have shown that the sweep schedule can be improved by increasing the number sweeps in the critical region. Annealing schedule optimizations alone have accounted for nearly an order of magnitude improvement over previous versions of the algorithm. In the context of spin glasses, where there is a broad distribution of computational hardness, tailoring the population size to the difficulty of the disorder realization yields the single largest improvement in efficiency. In addition, the sample-dependent population size results in simulations with both less work and higher overall accuracy. 

We still lack a theoretical understanding of the best sweep schedule.  Intuitively, we would like the Markov chain Monte Carlo subroutine to fully equilibrate each replica in the high temperature regime. However, in the low temperature regime, this is not feasible and the goal is to  equilibrate replicas only within their free energy minima, leaving  re-sampling  to properly redistribute replicas between distinct free energy minima. We have not yet found a principled way to achieve this goal. 

\appendix
\section{Covariance Inequality}\label{appendix_cov}
In Sec.\ \ref{section_theory_1} we showed that $\rho_t-1> \rho_f$, at least for the exact free energy version of PA, if the following inequality holds,
\begin{align}
&2\ \mathrm{cov} \left( \delta \tilde{R}_{k}, \sum_{j = N_{T-1}}^{k + 1} \sum_{i = 1}^{\tilde{R}_{j+1}} z(\tau_i^{j}) \right) \nonumber \\
&\qquad \qquad -\mathrm{var} \left( \sum_{j = N_{T-1}}^{k + 1} \sum_{i = 1}^{\tilde{R}_{j+1}} z(\tau_i^{j}) \right) >0. \label{eq:appcov}
\end{align} 
To establish this inequality, we begin by noting that $\delta \tilde{R}_k$  is the sum of deviations from the initial population size $R$ that have accumulated during each resampling step,
\begin{align}
\delta \tilde{R}_k &= \sum_{j = N_{T-1}}^{k + 1} \sum_{i = 1}^{\tilde{R}_{j+1}} \left( n(\tau_i^{j}) - 1 \right)\nonumber\\
&= \sum_{j = N_{T-1}}^{k + 1} \sum_{i = 1}^{\tilde{R}_{j+1}} \left( \tau_i^j + z(\tau_i^j) - 1 \right).
\end{align}
Using this result to expand the covariance term shows that the desired inequality,  Eq.\ (\ref{eq:appcov}) can be re-written as
\begin{align}
\label{eq:covin}
2\ \mathrm{cov} &\left( \sum_{j = N_{T-1}}^{k + 1} \sum_{i = 1}^{\tilde{R}_{j+1}} \tau_i^j, \, \sum_{j = N_{T-1}}^{k + 1} \sum_{i = 1}^{\tilde{R}_{j+1}} z(\tau_i^j) \right) \nonumber \\
+\ \mathrm{var} &\left( \sum_{j = N_{T-1}}^{k + 1} \sum_{i = 1}^{\tilde{R}_{j+1}} z(\tau_i^j) \right) >0 .
\end{align}
The variance is obviously non-negative but we do not have a proof that the covariance term is also non-negative.  However, we can motivate this assertion by noting that if the population at an earlier resampling step is stochastically increased ($z >0$) then later population sizes will tend to be increased, i.e.\ $\mathrm{cov}[\tilde{R}_{j'}, z(\tau_i^j) ] \geq 0$ for all $i$ and all $j'<j$. Furthermore, if the population is stochastically increased at an early resampling step, since it is now larger, it will better explore the low energy tail of the Gibbs distribution so that   $\mathrm{cov}[\tau_{i'}^{j'}, z(\tau_i^j) ] \geq 0$ for all $i$ and $i'$, and all $k<j$.  These two mechanisms both cause the covariance term in Eq.\ (\ref{eq:covin}) to be positive.   It is worth noting that $\mathrm{cov}[\tau_{i'}^{j'}, z(\tau_i^j)] = 0$ for all $i$ and $i'$, and all $j'<j$.
\section{Variance Expansion}\label{appendix_var}
The variance term from Eq.\ (\ref{eq:covin}) can be further expanded to get to a form similar to that of Eq. (\ref{eq:rhofratiorhot}). To do this, we make the approximation
\begin{equation}
\mathrm{var}\left( \sum_{j = N_{T-1}}^{k + 1} \sum_{i = 1}^{\tilde{R}_{j+1}} z(\tau_i^j) \right) \approx  \sum_{j = N_{T-1}}^{k + 1} \sum_{i = 1}^{\tilde{R}_{j+1}} \mathrm{var}[z(\tau_i^j)].
\end{equation}
It is possible to do this step exactly by noting that only $z_i$ from the same family can be correlated and by including covariance terms between intra-family $z(\tau_i)$, however, these terms have been found to be numerically insignificant.

The sum of the variances of $z(\tau_i)$ during an annealing step can be calculated using a method similar to that in Sec.\ \ref{section_theory_2}. We begin by writing the values and probabilities of $z(\tau)$,
\begin{equation}
  z(\tau) = \left\{
     \begin{array}{@{}l@{\thinspace}l}
       \tau - \lfloor\tau\rfloor  &\quad \text{w/ prob.}\ \lceil\tau\rceil - \tau \\
       \tau - \lceil\tau\rceil  &\quad \text{w/ prob.}\ \tau - \lfloor\tau\rfloor, \\
     \end{array}
   \right.
\end{equation}
which means that we can write the variance explicitly,
\begin{equation}
\mathrm{var}[z(\tau)] = (\tau - \lfloor\tau\rfloor)^2 (\lceil\tau\rceil - \tau) + (\tau - \lceil\tau\rceil)^2 (\tau - \lfloor\tau\rfloor). \label{eq:vartau}
\end{equation}
For culling fraction small, we can make the approximation that within a single annealing step, $\tau$ is a Gaussian random variable. This allows us to replace the sum of variances with an integral that can be calculated explicitly. Using Eq.\ (\ref{eq:vartau}),
\begin{align}
\sum_{i = 1}^{\tilde{R}_{j+1}} \mathrm{var}[z(\tau_i^j)] &\approx \int_{-\infty}^{\infty} \mathrm{var}(z(\tau))\mathscr{N}(\tau;1,\Delta\beta\sigma_E) \,\mathrm{d}\tau. \\
 &= 2\epsilon - 2\pi\epsilon^2.
\end{align}
Summing over all annealing steps and simplifying the expression assuming a constant culling fraction gives the desired result
\begin{align}
\sum_{j = N_{T-1}}^{k + 1} \sum_{i = 1}^{\tilde{R}_{j+1}} \mathrm{var}[z(\tau_i^j)] &\approx \sum_{j = N_{T-1}}^{k + 1} \left[ 2\epsilon(l) - 2\pi\epsilon(l)^2 \right]\\
&= (2\epsilon - 2\pi\epsilon^2)\,k.
\end{align}

\acknowledgments
This work was supported by the National Science Foundation (Grant No.~DMR-1507506).  We thank Sid Redner, Helmut Katzgraber, Wenlong Wang and Martin Weigel for useful discussions. 
\bibliographystyle{unsrtnat}
\bibliography{refs,references}
\end{document}